# THEORETICAL INVESTIGATION ON ELECTRONIC PROPERTIES OF TOPOLOGICAL MATERIALS: OPTICAL EXCITATIONS IN MÖBIUS CONJUGATED POLYMERS

Kikuo Harigaya

*Nanotechnology Research Institute, AIST, Tsukuba 305-8568, Japan*
*Synthetic Nano-Function Materials Project, AIST, Tsukuba 305-8568, Japan*

**Abstract**

Electronic structures and optical excitations in Möbius conjugated polymers are studied theoretically. Periodic and Möbius boundary conditions are applied to the tight binding model of poly(*para*-phenylene), taking into account of the exciton effects by long-range Coulomb interactions. We first discuss that oligomers with a few structural units are more effective than polymers, in order to measure effects of discrete wave numbers which are shift by the Möbius boundary from those of the periodic boundary. Next, calculations of the optical absorption spectra are reported. Certain components of the optical absorption for the electric field perpendicular to the polymer axis mix with the absorption spectra for the electric filed parallel with the polymer axis. Therefore, the polarization dependences of electric field of light can detect whether conjugated polymers have the Möbius boundary or not.

## 1. Introduction

Recently, low dimensional materials with peculiar boundary conditions, *i.e.*, Möbius boundaries have been synthesized: $NbSe_3$ [1] and aromatic hydrocarbons [2]. The Möbius strip consists of one surface which does not have the difference between the outer and inner surfaces. The orbitals of electrons are twisted during the travel along with the strip axis, and electronic states can be treated with the anti-periodic boundary conditions mathematically. Even though the presence of twisted π-electron systems has been predicted theoretically about forty years ago [3], structural perturbations by the topological characters might result in physical properties, so intensive investigations are being performed experimentally and theoretically.

In theoretical viewpoint, we have studied boundary condition effects in nanographite systems where carbon atoms are arrayed in one dimensional shape with zigzag edges [4-6]. Due to the presence of the Möbius boundary, magnetic domain wall states and helical magnetic orders are realized in spin alignments, and the domain wall appears in the charge density wave states, too. Such the abundant properties have been observed by their unique magnetic properties experimentally [7,8] for example.

In this paper, we investigate another candidate of twisted π-electron systems: conjugated polymers. We choose poly(*para*-phenylene) as a model material of conjugated polymers. Optical excitations in periodic systems have been studied using the tight binding model with long-range Coulomb interactions previously [9]. Exciton effects have been taken into account by the configuration interaction method. Here, we study optical excitations in Möbius poly(*para*-phenylene) strips.

## 2. Periodic Boundary Case

First we discuss optical absorption of poly(*para*-phenylene) for the periodic boundary. In the model [9], the Ohno potential is used with the parameters, the onsite repulsion $U=2.5t$ and the long range component $V=1.3t$, $t$ being the hopping integral of the neighboring carbon atoms. The number of phenyl rings is $N=20$, and the ring torsion angle is $\Phi=0°$. Figure 1 shows the calculated absorption with the Lorentzian broadening $\gamma=0.15t$. The circle of the polymer axis is assumed to be placed within the x-y plane. The z-axis is perpendicular to the polymer axis. When the electric field of light is parallel with the polymer axis [Fig. 1 (a)], the lowest optical excitations appear at the energy about $1.4t$. They are excitations between the extended states. The higher feature at $2.4t$ is due to the optical excitations between the localized occupied and unoccupied orbitals. When the electric field is perpendicular to the polymer axis [Fig. 1 (b)], the optical excitations appear at the energies larger than about $2.2t$. They are excitations between the extended states and the localized orbitals. Figure 1 (c) shows the averaged spectra with respect to the polarization of light, *i.e.*, the sum over x-, y, and z-polarizations. The overall features have been observed experimentally. For details, refer to the published report [9].

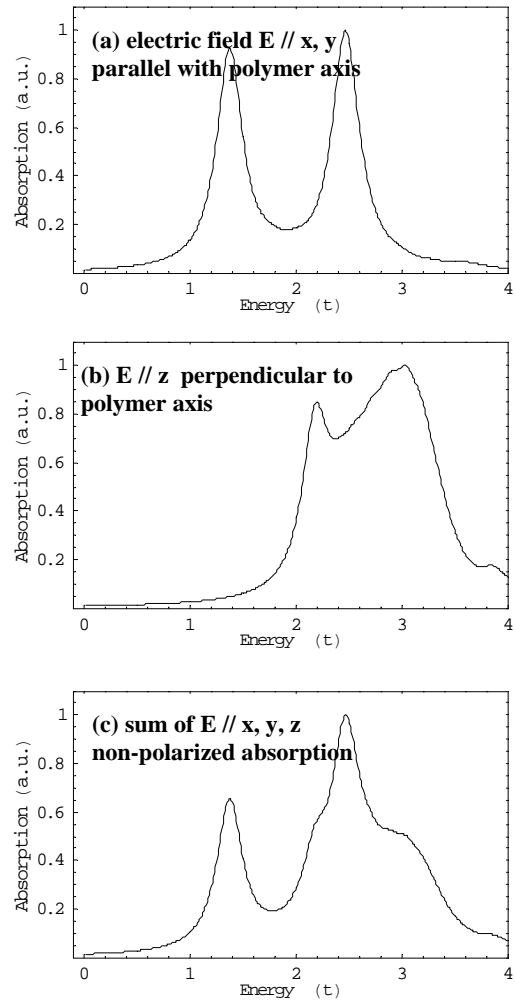

Fig. 1. Optical absorption spectra for the periodic boundary case: (a) electric field of light parallel with the polymer axis, (b) electric field perpendicular to the polymer axis, and (c) non-polarized absorption.

## 3. Möbius Boundary Case

The difference between the periodic and Möbius boundary conditions is regarded as coming from addition of the anti-periodicity. As well known in the textbook of condensed matter physics, the allowed wave numbers are different in periodic and anti-periodic boundary conditions. Because the allowed states in the wave number space are populated densely in the long enough polymer, the effects of the boundary condition difference are too small to be observed. However, in oligomers with a few phenyl rings, the allowed states are so sparsely distributed that the boundary condition effects could be measured. The recent synthesis of Möbius aromatic systems [2] will promote synthesis of Möbius polymers in view of the presence of many kinds of ring polymers [10,11].

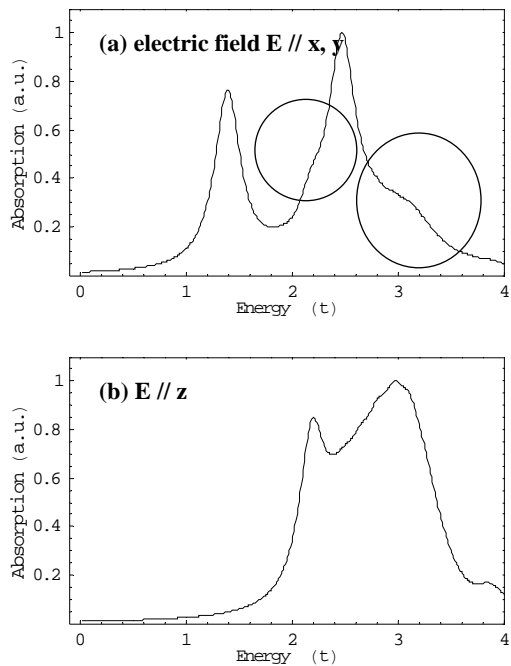

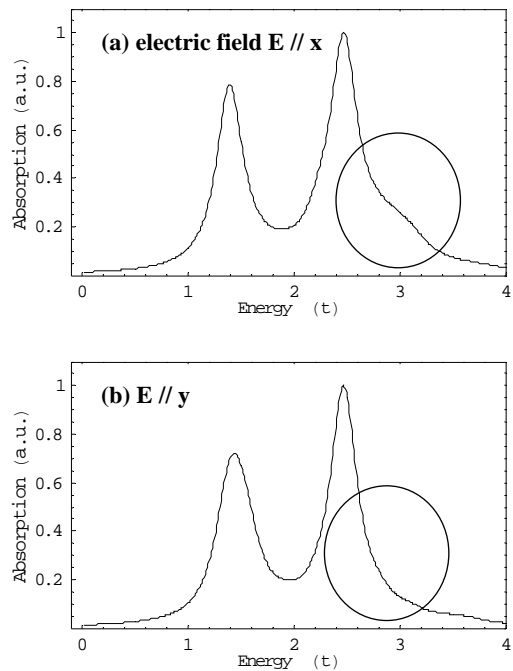

Fig. 2. Optical absorption spectra for the Möbius boundary case I: (a) electric field of light parallel with the polymer axis, (b) electric field perpendicular to the polymer axis. The circles show features coming from the mix of perpendicular polarization.

Fig. 3. Optical absorption spectra for the Möbius boundary case II: (a) electric field of light parallel with the x-axis, and (b) y-axis. The circles show features coming from the mix of perpendicular polarization.

Now, we consider two cases of Möbius boundaries. The first case (I) is that the ring torsions occur uniformly over the polymer with the torsion angle $\Phi=180°/N$. The phenyl rings rotate helically along with the polymer. The polymer axis is represented by a circle in the x-y plane as in the periodic boundary case. The second case (II) is that the twist due to the Möbius boundary is localized among five phenyls where the polymer circle crosses the x-axis. The twist angle from the torsion is taken as $\Phi=30°$. Hereafter, the calculated absorption spectra are shown for the number of phenyl rings N=20.

In the case I, the plane which includes the polymer is almost parallel with the x-y plane in a certain part of the polymer. Among this part, almost perpendicular polarization is realized, and therefore mixing of the absorption coming from perpendicular polarization [Fig. 1 (b)] is expected. Figure 2 (a) shows the actual calculation. There are two weak features among the energies 2.2t and 3.0t owing to the mixing. There is no difference between the x- and y-polarizations, because the system is uniform. On the other hand, the absorption of the perpendicular polarization does not show any polarization dependence [Fig. 2 (b)].

In the case II, the polymer plane can become parallel with the x-y plane at the region where the twist is present. There is nearly perpendicular polarization among this region. The mixing of the perpendicular polarization occurs among the energies 3.0t as shown in Fig. 3 (a), where the electric field of light is parallel with the x-axis. For the case that the field is along the y-axis, the mixing of the perpendicular polarization is very weak because of the localization of the twist [Fig. 3 (b)].

## 4. Summary

Electronic structures and optical excitations in poly(*para*-phenylene) with periodic and Möbius boundaries have been studied by taking into account of the exciton effects. We have discussed that oligomers with a few structural units are more effective than polymers, in order to measure effects of discrete wave numbers which are shift by the Möbius boundary from those of the periodic boundary. In the calculated optical absorption spectra, certain components of the optical absorption for the electric field perpendicular to the polymer axis mix with the absorption spectra for the electric filed parallel with the polymer axis. The polarization dependences of electric field of light can detect whether conjugated polymers have the Möbius boundary or not.

### Acknowledgments

This work has been supported partly by Special Coordination Funds for Promoting Science and Technology, and by NEDO under the Nanotechnology Program.

### References

[1] S. Tanda *et al.*, Nature 417 (2002) 397.
[2] D. Ajami *et al.*, Nature 426 (2003) 819.
[3] E. Heilbronner, Tetrahedron Lett. (1964) 1923.
[4] K. Wakabayashi and K. Harigaya, J. Phys. Soc. Jpn. 72 (2003) 998.
[5] A. Yamashiro *et al.*, Phys. Rev. B 68 (2003) 193410.
[6] A. Yamashiro *et al.*, Physica E 22 (2004) 688.
[7] Y. Shibayama *et al.*, Phys. Rev. Lett. 84 (2000) 1744.
[8] H. Sato *et al.*, Solid State Commun. 125 (2003) 641.
[9] K. Harigaya, J. Phys.: Condens. Matter 10 (1998) 7679.
[10] E. Mena-Osteritz, Adv. Mater. 14 (2002) 609.
[11] M. Mayor and C. Didschies, Angew. Chem. Int. Ed. 42 (2003) 3176.